\documentclass[aps,floatfix,twocolumn]{revtex4}
\usepackage{graphics,epsfig}
\usepackage{graphicx}
\begin{document}

\title{Transforming the Einstein static Universe into physically acceptable static fluid spheres}
\author{Kayll Lake \cite{email}}
\affiliation{Department of Physics, Queen's University, Kingston,
Ontario, Canada, K7L 3N6 }
\date{\today}

\begin{abstract}
The staid subject of exact static spherically symmetric perfect fluid solutions of Einstein's equations has been reinvigorated in the last decade. We now have several solution generating techniques which give rise to new exact solutions. Here the Einstein static Universe is transformed into a physically acceptable solution the properties of which are examined in detail. The emphasis here is on the importance of the integration constants that these generating techniques introduce.
\end{abstract}
\maketitle
\section{Introduction}
A decade ago, as a demonstration of the use of computer algebra \cite{dellake}, it was shown that few of the alleged exact static spherically symmetric perfect fluid solutions of Einstein's equations were in fact correct and even fewer made physical sense \cite{physical}. Since then this field of study has been reinvigorated with the development of several solution generating techniques which give rise to new exact solutions. These techniques are based on the pressure isotropy condition, either looked at as a differential equation, following Wyman \cite{wyman}, or by way of invariance properties,  following Buchdahl \cite{buchdahl}. For example, following the work of Rahman and Visser \cite{visser}, I showed that several of the acceptable solutions given in \cite{dellake} follow from a simple algorithm \cite{lake} which in fact generates an infinite number of physically acceptable solutions. Since then several works of interest have appeared: \cite{martin}, \cite{petarpa1}, \cite{petarpa2},  \cite{petarpa3},  \cite{petarpa4}, \cite{petarpa5} and \cite{petarpa6}. It is remarkable that such a staid old subject has bounced back to life.

Here I do not add to these generating techniques but rather make use of one of them to do what they are intended to do: generate a new physically interesting exact solution of Einstein's equations. The example provided serves to emphasize  the importance of the integration constant that the generating technique introduces.
\section{Generating technique}
Every spacetime $\mathcal{M}$ with metric \cite{notation}
\begin{equation}
ds^2_{\mathcal{M}}=\frac{dr^2}{1-\frac{2m(r)}{r}}+r^2d\Omega^2-e^{2(\Phi(r)+\xi(r))}dt^2,
\label{standardform}
\end{equation}
where $d\Omega^2 \equiv d\theta^2+\sin(\theta)^2 d\phi^2$,
is an exact perfect fluid solution of Einstein's equations as long as
\begin{equation}
m =\frac{\int \!b ( r) {e^{\int \!a ( r)
{dr}}}{dr}+\mathcal{C}_{1}}{{e^{\int \!a(r) {dr}}}} \label{mass}
\end{equation}
and
\begin{equation}
\xi=\ln(\mathcal{C}_{2} \int e^{-A(r)} dr+\mathcal{C}_{3}) \label{xi}
\end{equation}
where
\begin{equation}
A\equiv\int\frac{c(r)dr}{r\Phi^{'}+1}-\int\frac{dr}{(r\Phi^{'}+1)(1-\frac{2m}{r})r} \label{A},
\end{equation}
$^{'}\equiv\frac{d}{dr}$ and the $\mathcal{C}_{n}$ are constants.
The functions $a(r), b(r)$ and $c(r)$ are given by
\begin{equation}
a \equiv{\frac {2 r^{2}(\Phi^{''}+ \Phi ^{'\;2})
 -3 (r\,  \Phi^{'}
 \, +1)}{r (  r \Phi^{'}
  +1 ) }},\label{a}
\end{equation}

\begin{equation}
b \equiv{\frac {r ( r (\Phi^{''}   +  \Phi ^{'\;2}
) -\Phi^{'}   ) }{
 r \Phi^{'}    +1}},\label{b}
\end{equation}
and
\begin{equation}
c \equiv -r\Phi^{''}+r\Phi ^{'\;2}
+2\Phi^{'}.\label{c}
\end{equation}
The algorithm can be executed subject to the specification of the function $\Phi$ (as well as smoothness and boundary conditions \cite{lake}) and the constants $\mathcal{C}_{n}$ \cite{form}.

The procedure I consider here will assume $\mathcal{C}_{2}=0$ and so $\mathcal{C}_{3}$
is disposable (it can be absorbed into the scale of $t$). Call these spacetimes $\mathcal{N}$. Further, I will write the spacetimes in the form
\begin{equation}
ds^2_{\mathcal{O}}=e^{2\chi(r)}\left(\frac{dr^2}{1-\frac{2M(r)}{r}}+r^2d\Omega^2-e^{2\Phi(r)}dt^2\right)\label{newform}
\end{equation}
where $M$, as with $m$ in $\mathcal{M}$, is constructed so as to make $\mathcal{O}$ a perfect fluid. That is, for $M$, (\ref{a}) is replaced by
\begin{equation}
\tilde{a}=\frac{2r^2(\Phi^{''}+\Phi ^{'\;2})-3(r\Phi^{'}+1)+4r^2(\chi^{''}-\chi^{'\;2})-6r\chi^{'}}{r(r\Phi^{'}+1+2r\chi^{'})}\label{newa}
\end{equation}
and (\ref{b}) by
\begin{equation}
\tilde{b}=\frac{r(r(\Phi^{''}+\Phi ^{'\;2})-\Phi^{'}+2(r\chi^{''}-r\chi^{'\;2}-\chi^{'}))}{r\Phi^{'}+1+2r\chi^{'}}\label{newb}
\end{equation}
in (\ref{mass}). Unlike $m$ however, $M$ is no longer the effective gravitational mass \cite{lake}.

It is important to note that $\mathcal{O}$ is not a conformal transformation of $\mathcal{N}$ (due to the restrictions on $M$)  and it is no more general than $\mathcal{N}$ as it is merely a coordinate transformation of $\mathcal{N}$ \cite{transform}. We refer to the case $\chi=0$ as the ``seed" of the spacetimes $\mathcal{O}$. The usefulness of the form (\ref{newform}) derives from the fact that we can clearly recognize the seed.

\section{$\Phi = 0$}
The simplest seed for $\mathcal{O}$ is $\Phi=\mathcal{C}$ where $\mathcal{C}$ is a constant which, by choice of scale for $t$, we can set to zero. It follows that $m = \mathcal{C}_{1}r^3$ and the seed is simply the Einstein static Universe. (A cosmological constant $\Lambda=2 \mathcal{C}_{1}$ can be introduced to give zero pressure, but this is not done here \cite{bohmer}). Given this seed, the regularity conditions on $\chi$ are \cite{dellake}
\begin{equation}
|\chi(0)|<\infty,\;\;\;\;\chi^{'}(0)=0\label{regular}
\end{equation}
and so the simplest non-trivial form of $\chi$ satisfying these conditions is
\begin{equation}
\chi=\mathcal{C}_{4}+\mathcal{C}_{5}r^2,\;\;\label{simple}
\end{equation}
where $\mathcal{C}_{4}$ and $\mathcal{C}_{5}\neq0$ are constants \cite{kuch}. Since the constant $\mathcal{C}_{4}$ simply scales the energy density and pressure by $1/e^{2\mathcal{C}_{4}}$, without loss in physical generality we set $\mathcal{C}_{4}=0$. Since $\mathcal{C}_{5}$ can be absorbed into a redefinition of $r$ (and a rescaling of the as yet to be chosen constant $\mathcal{C}_{1}$) we set $\mathcal{C}_{5}=1$ so that without any loss in physical generality we take $\chi=r^2$. We now have
\begin{equation}
M=\frac{r^3}{R^2}\left(2+\frac{e^{2 r^2}}{R}(\mathcal{C}_{1}-\sqrt{2 \pi e} \;\texttt{erf}(\frac{R}{\sqrt{2}}))\right)
\label{MR}
\end{equation}
where $R \equiv \sqrt{1+4r^2}$ and $\texttt{erf}$ is the error function. Whereas we could of course continue our discussion in terms of the coordinates used in (\ref{newform}), we now revert to more traditional coordinates.

\begin{figure}[ht]
\epsfig{file=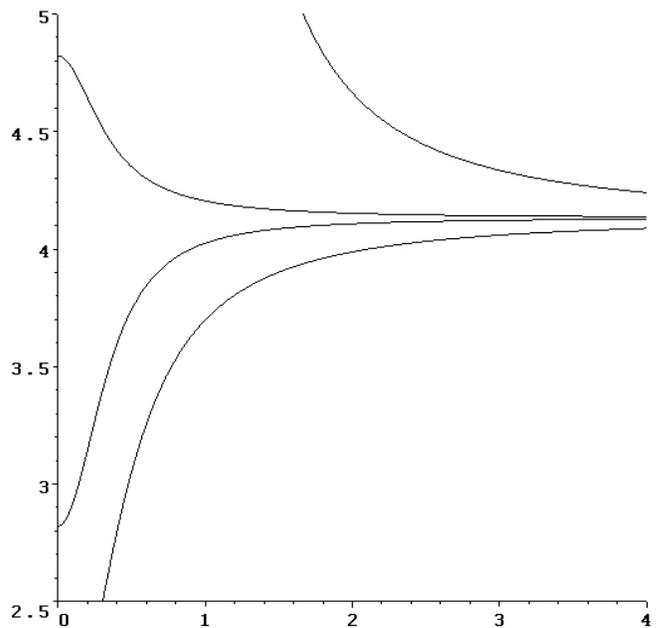,height=3.5in,width=3.5in,angle=0}
\caption{\label{gradients}The abscissa is $\textrm{r}$ and the ordinate is $\mathcal{C}_{1}$.
Top down the curves show $\rho^{'}=0$, $p=0$, $\rho=0$ and $p^{'}=0$.
The curves $p=0$ and $\rho=0$ intersect the ordinate at $2+\sqrt{2 \pi e}\;\texttt{erf}(\frac{1}{\sqrt{2}})$ and
$\sqrt{2 \pi e}\;\texttt{erf}(\frac{1}{\sqrt{2}})$ respectively. They are both asymptotic to $\mathcal{C}_{1}=\sqrt{2 \pi e}$.}
\end{figure}

\begin{figure}[ht]
\epsfig{file=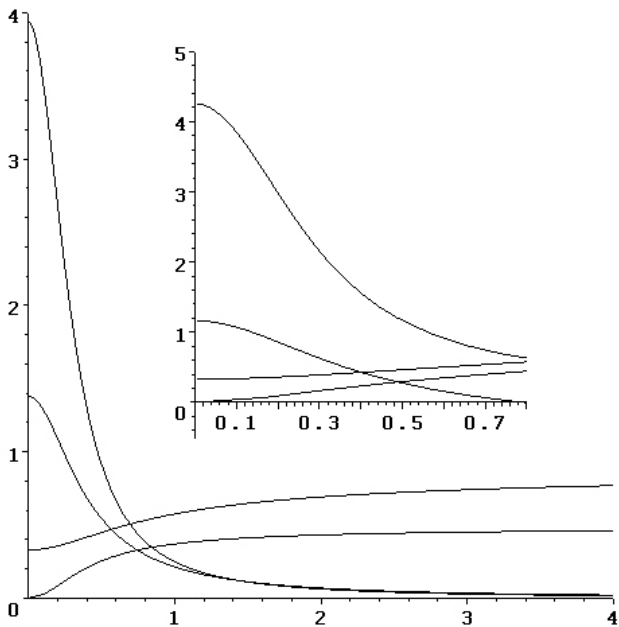,height=3.5in,width=3.5in,angle=0}
\caption{\label{properties}Some internal properties of the solutions.
Top down at the origin the curves give $\rho/2$, $p$, $v_{s}$ and $2\tilde{M}/\textrm{r}$ where $v_{s}$ is the adiabatic sound speed and $\tilde{M}$ the effective gravitational mass. The main image is for asymptotic case $\mathcal{C}_{1}=\sqrt{2 \pi e}$ and the insert is for $\mathcal{C}_{1}=4.24$ which terminates at $p=0$ where $\textrm{r} \simeq 0.7914$.}
\end{figure}

Under the coordinate transformation $e^{r^2}r=\textrm{r}$ we now have
\begin{equation}
ds^2=\mathcal{F}(\textrm{r})d\textrm{r}^2+\textrm{r}^2d\Omega^2-\frac{2\textrm{r}^2}{\mathcal{H}(\textrm{r})}dt^2\label{tansfr}
\end{equation}
where
\begin{equation}
\mathcal{F}=\frac{\mathcal{J}^{3}}{\left(\mathcal{J}+2\textrm{r}^2(\sqrt{2 \pi e} \; \mathcal{E}-\mathcal{C}_{1})\right)(1+\mathcal{H})^2}\label{F}
\end{equation}
with
\begin{equation}\label{J}
\mathcal{J} \equiv \sqrt{1+2\;\mathcal{H}},
\end{equation}
\begin{equation}
\mathcal{E} \equiv \texttt{erf}(\frac{\mathcal{J}}{\sqrt{2}})\label{lamberte}
\end{equation}
and
\begin{equation}
\mathcal{H} \equiv \mathcal{W}(2\textrm{r}^2)\label{lambert}
\end{equation}
where $\mathcal{W}$ is the Lambert W function \cite{lambert}. As shown in FIG.~\ref{gradients}, the constant $\mathcal{C}_{1}$ plays the central role regarding the physical acceptability of these solutions. For
\begin{equation}\label{unphysical}
\mathcal{C}_{1} < \sqrt{2 \pi e}
\end{equation}
$\rho$ vanishes while $p>0$ which is physically unacceptable. For
\begin{equation}\label{border}
\mathcal{C}_{1} = \sqrt{2 \pi e}
\end{equation}
the solution is global, not isolated, and $\rho$ and $p$ $\rightarrow 0$ only as $\textrm{r} \rightarrow \infty$.
Finally, for
\begin{equation}\label{physical}
 \sqrt{2 \pi e} <  \mathcal{C}_{1} < 2+\sqrt{2 \pi e}\;\texttt{erf}(\frac{1}{\sqrt{2}})
\end{equation}
the pressure $p$ vanishes at finite $\textrm{r}>0$ and the solutions match onto a vacuum exterior by way of continuity of the effective gravitational mass. The density contrast at the boundary increases as $\mathcal{C}_{1}$ increases.
Throughout the physically acceptable distributions $\rho$ and $p$ are monotone decreasing and the adiabatic sound speed $v_{s}$ is subluminal. Some properties are shown in FIG.~\ref{properties}.

We record here in explicit form the essential physical elements of the solutions:
The effective gravitational mass is given by
\begin{equation}\label{massf}
\tilde{M}=\frac{\textrm{r}}{2}\left(\frac{\mathcal{F}-1}{\mathcal{F}}\right)
\end{equation}
with $\mathcal{F}$ given by (\ref{F}), the energy density is given by
\begin{equation}\label{density}
8 \pi\;\rho={\frac {2\mathcal{K}\textrm{r}^{2} \left( \mathcal{C}_{1}-\sqrt {2 \pi e}\;\mathcal{E}
 \right) -3\,\mathcal{L}}{4\mathcal{J}^{5}\textrm{r}^{2}}}
\end{equation}
where
\begin{equation}\label{K}
\mathcal{K} \equiv 3\mathcal{J}^6+8\mathcal{J}^4-5\mathcal{J}^2+6
\end{equation}
and
\begin{equation}\label{l}
\mathcal{L} \equiv (\mathcal{J}^2+2)(\mathcal{J}-1)^2(\mathcal{J}+1)^2\mathcal{J},
\end{equation}
and the isotropic pressure is given by
\begin{equation}\label{pressure}
8 \pi\;p=\frac {2\mathcal{P}\textrm{r}^{2} \left(\sqrt {2 \pi e}\;\mathcal{E}- \mathcal{C}_{1}
 \right) +\mathcal{Q}}{4\mathcal{J}^{3}\textrm{r}^{2}}
\end{equation}
where
\begin{equation}\label{P}
\mathcal{P} \equiv (3\mathcal{J}^2-1)(\mathcal{J}^2+1)
\end{equation}
and
\begin{equation}\label{Q}
\mathcal{Q} \equiv (3\mathcal{J}^2+1)(\mathcal{J}-1)(\mathcal{J}+1)\mathcal{J}.
\end{equation}
Finally, the square of the adiabatic sound speed is given by
\begin{equation}\label{sound}
v_{s}^2=\mathcal{J}^2\frac {2\mathcal{R}\textrm{r}^{2} \left(\sqrt {2 \pi e}\;\mathcal{E}- \mathcal{C}_{1}
 \right) +\mathcal{S}}{2\mathcal{T}\textrm{r}^{2} \left(\mathcal{C}_{1}-\sqrt {2 \pi e}\;\mathcal{E}
 \right) +\mathcal{U}}
\end{equation}
where
\begin{equation}\label{R}
\mathcal{R} \equiv (\mathcal{J}^2-3)(\mathcal{J}-1)(\mathcal{J}+1)\mathcal{J},
\end{equation}

\begin{equation}\label{S}
\mathcal{S} \equiv 3\mathcal{J}^4-2\mathcal{J}^2+3,
\end{equation}

\begin{equation}\label{T}
\mathcal{T} \equiv (\mathcal{J}^4-\mathcal{J}^2+6)(\mathcal{J}-1)(\mathcal{J}+1)\mathcal{J},
\end{equation}
and
\begin{equation}\label{U}
\mathcal{U} \equiv -3\mathcal{J}^6+8\mathcal{J}^4-15\mathcal{J}^2+30.
\end{equation}

\bigskip

\section{Discussion}

The Einstein static Universe has been transformed into a class of physically acceptable static fluid spheres whose physical properties have been written out in explicit form. The technique is but a coordinate transformation of one discussed previously \cite{lake} but it allows a clear understanding how various spacetimes can be interrelated \cite{others}.
\begin{acknowledgments}
This work was supported by a grant from the Natural Sciences and
Engineering Research Council of Canada. Portions of this work were
made possible by use of \textit{GRTensorII} \cite{grt}.
\end{acknowledgments}


\begin{thebibliography}{}
\bibitem{physical}The conditions used in \cite{dellake} were: (i) isotropy of the
pressure (otherwise any metric is a ``solution"), (ii) regularity
of associated invariants at the origin, (iii) positivity of the pressure and energy
density, (iv) vanishing of the pressure at a finite
boundary, (v) monotone decrease of the pressure and energy density to the
boundary and (vi) subluminal adiabatic sound speed. In addition to
these, a monotone decrease in the subluminal adiabatic sound speed
was considered desirable. Probably only issues regarding the ``sound speed" would offer room for debate.
\bibitem{form} Up to notation and a coordinate transformation the algorithm given is equivalent to Theorems 1 and 2 in \cite{petarpa6}.
\bibitem{transform}Define $re^{\chi(r)}\equiv \textrm{r}$ and $\Phi(r)+\chi(r)\equiv \psi(\textrm{r})$ and write $g_{_{\textrm{r}\textrm{r}}}=1/(1-2\tilde{m}(\textrm{r})/\textrm{r})$. Then $\tilde{m}(\textrm{r})$ is given by (\ref{mass}) with $\Phi$ replaced by $\psi$ and $r$ by $\textrm{r}$.
\bibitem{kuch}This solution is distinct from but in a sense complimentary to the solution generated by $\chi=0$ and $\Phi=\mathcal{C}_{4}+\mathcal{C}_{5}r^2$; Kuch2 III in the notation of \cite{dellake}.
\bibitem{others}Closed form exact solutions have also been found for other choices of $\Phi$ and $\chi$ by Alex Klotz, C\'{e}dric Grenon and Pascal Elahi.
\end{thebibliography}
\end{document}